\documentstyle[12pt,epsfig]{article}
\begin{document}
\begin{center}
{\Large\bf Baryon charge transfer and production asymmetry of 
$\Lambda^0$/$\bar{\Lambda^0}$ in hadron interactions.}
\vspace{10mm}

 O.I.Piskounova\\

 P.N.Lebedev Physical Institute of Russian Academy of Science,
Moscow \\
 Russia \\
\end{center}
\vspace{12mm}
\begin{abstract}
The predictions were done for asymmetry between production spectra of $\Lambda^0$ 
and $\bar{\Lambda^0}$ at the energy of LHC experiments. The value of A(s) should be
situated in the band between two curves that are calculated in Quark-Gluon 
String Model with two possible values of intercept $\alpha_{SJ}$(0)=0,5 and 0,9. Both curves
describe the asymmetries measured at lower energies up to RHIC experiments.
The data of H1 experiment can be fitted only with $\alpha_{SJ}$(0)=0,9. 

\end{abstract}
\newpage

\section{Introduction to the QGSM}

The phenomena of nonzero asymmetry of baryon production with nonbaryonic beams 
($\pi$,$\mu$,e) was mentioned and explained in few theoretical papers. Baryon
charge can be transfered from proton or nucleus target through the large
rapidity gap with the string junction from baryonic target. In baryonic beam 
reactions (p,A,etc.)this effect is displayed in the valuable baryon/antibaryon spectrum asymmetry 
in the central region (y=0).

Every theoretical discourse on baryon charge transfer appeals to the value of  
intercept, $\alpha_{SJ}$(0), that is the intercept of Regge-trajectory of imaginary 
particles which consist only of string junctions from baryon and antibaryon.
Practically, the models that can account this effect are only non perturbative
QCD phenomenological models: Dual Parton Model (DPM) \cite{DPM} and 
Qark Gluon String Model (QGSM) \cite{QGSM} as well as DPMJET 
Monte Carlo expantion of previous two models \cite{DPMJ} .
Both analytical models are similar and were based on the same Regge
asymthotical presentation of constituent quark structure functions and string
(quark) fragmentation functions. Here we are considering QGSM. 
In the comparison to the other models, QGSM accounts many Pomeron
exchanges. This advantage works very well to give us the correct description of
particle production cross sections at very high energies.
The QGSM procedures of constructing of quark/diquark structure functions and
fragmentation functions were presented in many previous publications 
\cite{QGSM,kaidalov,oldbaryons}.
We take into consideration the $\pi-p$ interaction that gives the same asymmetries as
$\gamma-p$ reaction. The spectra in this reaction are more sensitive to baryon excess
in the region of positive $x_F$ than spectra of baryons in $p-p$ collisions.

\section{Diquark Fragmentation Function and String Junction Transfer}

Three cases of fragmentation of diquark of target proton are shown in Fig.1.
The fragmentation functions of diquark and quark chains into strange baryons 
or antibaryons are based on the rules written in \cite{kaidalov}.

The $uu$- and $ud$-diquark fragmentation function that corresponds to the diagram Fig.1 a) 
includes the constant $a_f^{\Lambda^0}$ 
which could be interpreted as "leading" parameter, but the value of $a_f^{\Lambda^0}$ 
is fixed due to the baryon number sum rule and should be approximately equal to the value taken for 
$\Lambda^0$ spectra in our previous calculations \cite{oldbaryons}:

\begin{equation}
{\cal D}_{dd}^{\Lambda^0}(z)=\frac{a_f^{\Lambda^0}}{a_0^{\bar{\Lambda}^0}z}
z^{2\alpha_R(0)-2\alpha_N(0)}
(1-z)^{-\alpha_{\phi}(0)+\lambda+2(1-\alpha_R(0))},
\end{equation}
where the term $z^{2\alpha_R(0)-2\alpha_N(0)}$ means the probability for
initial diquark to have z close to 0; the intercepts of Regge trajectories,
$\alpha_R(0)$, $\alpha_N(0)$ and $\alpha_{\phi}(0)$ are taken of the same 
values as in \cite{oldbaryons}, 0.5, -0.5 and 0. correspondingly. The
$\lambda$ parameter is an remnant of transverse momenta dependence, it is equal to 0,5
here ( for more information see the early publications \cite{QGSM,oldbaryons}).

The production of stange baryons in the central rapidity region as it is shown in Fig.1.b) 
will not contribute to the asymmetry because this case of fragmentation gives the equal amounts of  $\Lambda^0$ and $\bar{\Lambda^0}$ with the density $a_0^{\bar{\Lambda^0}}$ per the unit of rapidity.

It is important here to keep in mind the possibility to create $\Lambda^0$
baryon only on the  basis of  string junction taken from interacting target proton, see Fig.1c). 
The string junction  brings 
the positive baryon number in baryons and the negative one in antibaryons. So only positive baryon number can be transfered from proton. 
The fragmentation function of string junction that can be brought to the region $z>0$
is of the similar form as diquark FF written above :
  
\begin{equation}
{\cal D}_{SJ}^{\Lambda^0}(z)=\frac{a_f^{\Lambda^0}}{a_0^{\bar{\Lambda}_c}z}
z^{1-\alpha_{SJ}(0)}(1-z)^{-\alpha_{\phi}(0)+\lambda+2(1-\alpha_R(0))},
\end{equation}
where $\alpha_{SJ}(0)$ is the intercept of string junction Regge trajectory. 
We are discussing here two possible values 
of string junction intercept: 0,5 \cite{strjunction} and 1,0 \cite{kopeliovich}. 
It becomes significant when we study the baryon spectra in pion-proton or photon-proton 
interactions at high energies. 

\section{Comparison with Experimental Data}

Asymmetry A(y) between spectra of baryons and antibaryons is defined as:

\begin{equation}
A(y)=\frac{dN^{\Lambda^0}/dy-dN^{\bar{\Lambda^0}}/dy}
{dN^{\Lambda^0}/dy+dN^{\bar{\Lambda^0}}/dy},
\end{equation}

What evidence we have for nonzero baryon production asymmetry? Ratio between baryons 
and antibaryons was measured in few experiments for proton-proton fixed target interactions (LEBS-EHS \cite{EHS}, NA49 \cite{NA49}). These measurements gave the value of asymmetry in the central region (y=0) that is equal 0.5 - 0.3. 
The ISR data was withdrown from two experiments \cite{ISR}, so the error bars are mostly due to
the big difference between their results. In pion-proton interactions data from E769 \cite{E769} we can see the y dependence of asymmetry and the measured asymmetry was multiplied by 2 in order to compare with pp asymmetry.

The data from these experiments can be presented in one plot, A($\Delta$y), where $\Delta$y is the rapidity distance from interacting target-proton (see Fig.2). It is 
seen that the points are situated on the stright line. If we add the data from proton-nucleus experiments (HERA-b \cite{herab} and RHIC \cite{rhic}), points are still approximately on one
line. Only the STAR asymmetry point at sqrt(s)=130 GeV can be interpreted as a sign
that the curve is bent. And the result of H1 experiment at HERA \cite{H1} calls
certainly for steeper line.

What means we have in QGSM to describe this dependence? In the paper \cite{arakelyan}
 it was shown that the data of E769 and H1 experiments can not be described with the same 
value of $\alpha_{SJ}$(0): the points at lower energy are corresponding to $\alpha_{SJ}$(0)=0,5, but H1 point requires $\alpha_{SJ}$(0)=0,9.

\section{Summary}

My intension in this paper was to show the band of production asymmetries of 
$\Lambda^0$ and $\bar{\Lambda^0}$ that can be predicted 
for LHC experiments between two possibilities announced above for $\alpha_{SJ}$(0).
The results are shown in Fig.2. Solid line represents the case of $\alpha_{SJ}$(0)=0,9. This 
curve fits the data at low energies (small $\Delta$y) due to varying the energy splitting 
between string junction and diquark configuration: 0,1*SJ+0,9*DQ. What we had to tune 
also was the fragmentation parameter af=0,15 instead of 0,2 accepted in previous 
papers. 
There are also the curve for $\alpha_{SJ}$(0)=0,5 shown in Fig.2 with dashed line. This 
line certainly doesn't fit the H1 point and gives negligible asymmetry at the 
energy of LHC experiment. Finally, we have the prediction for strange baryon asymmetries 
at LHC within the range: $0,003 < A_{\Lambda^0} < 0,04$.
The same procedure have to be applied to the production of charmed baryon to get the predictions 
for $\Lambda_c$/$\bar{\Lambda_c}$ asymmery at LHC energy.

\section{Acknowlegments}

Author expresses her gratitude to Prof. Alexei Kaidalov, Dr. Dmitri Ozerov and Dr. Yuli Shabelski
for the discussions on theory and providing her with experimental data. This work was supported from
RFBR-DFG grant 04-02-04026a.

\newpage
\begin{figure}[th]
\centerline{\epsfig{figure=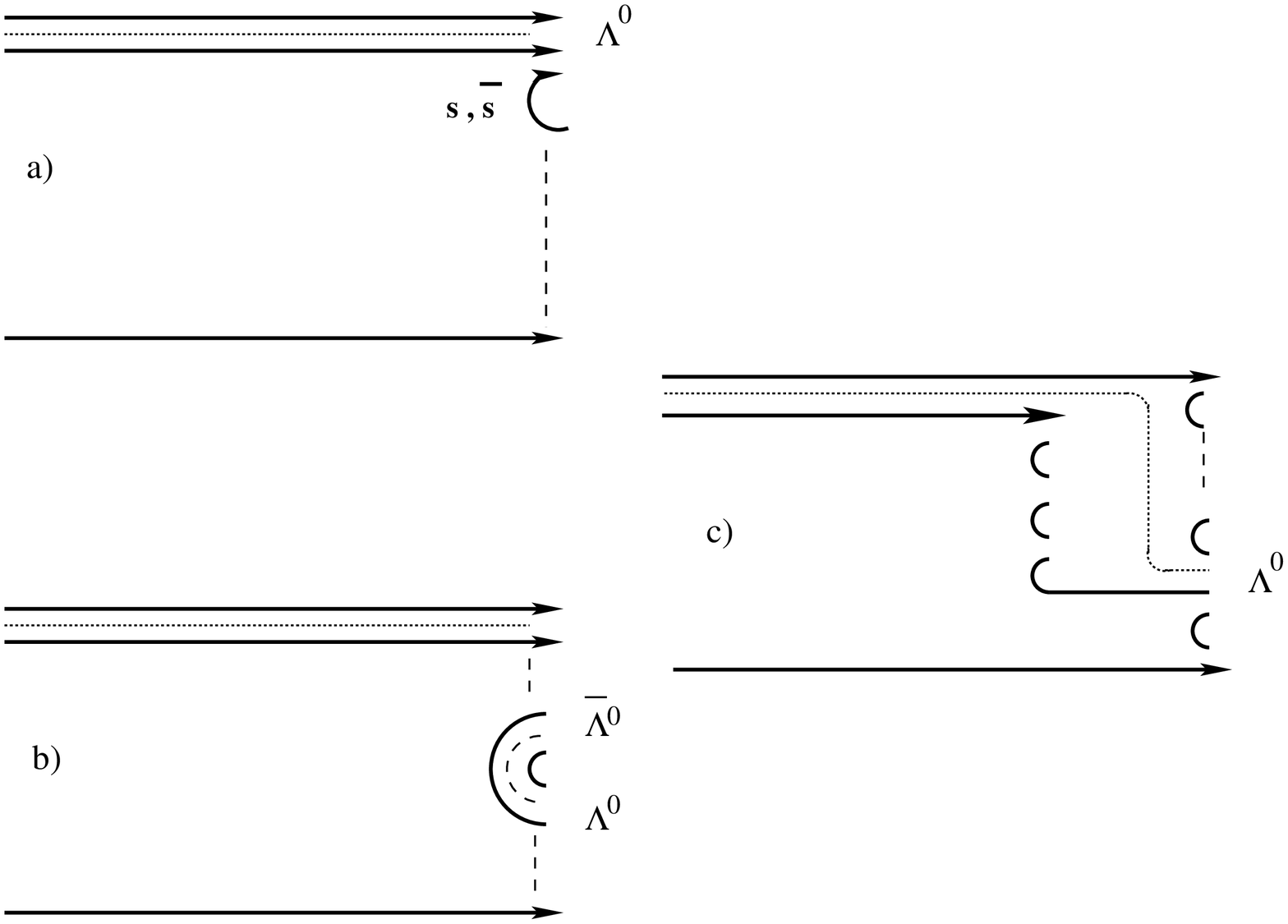,height=12cm,width=10cm}}
\caption{\label{Fragm_diagrams} The fragmentation of diquark chain: a) leading fragmentation
into $\Lambda^0$, b) central fragmentation into $\bar{\Lambda^0}$ and $\Lambda^0$ , c) fragmentation by string junction.}
\end{figure}

\newpage
\begin{figure}[th]
\centerline{\epsfig{figure=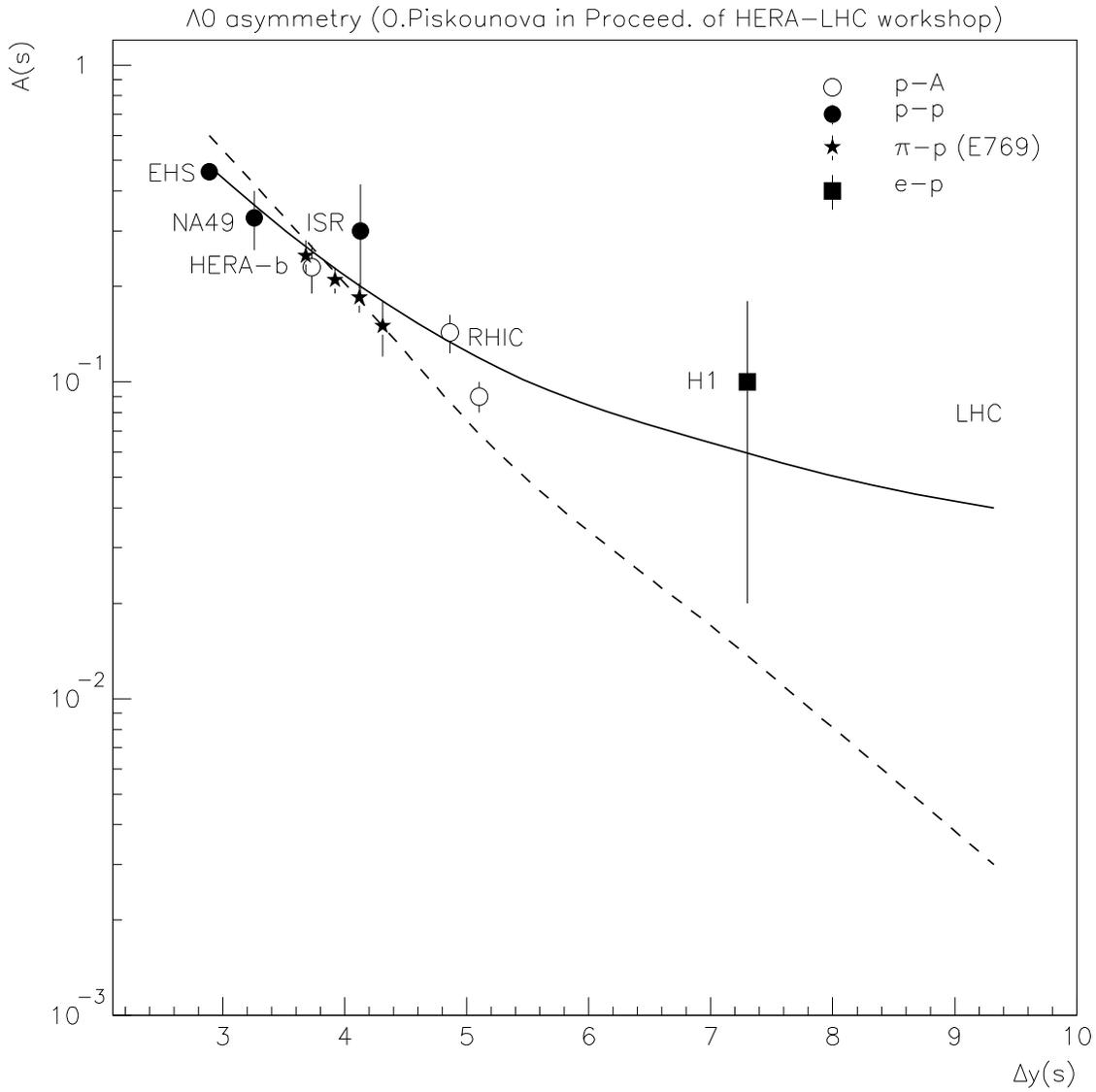,height=16cm,width=16cm}}
\caption{\label{Lambda0_asym} Asymmetry in $\Lambda^0$ 
and $\bar{\Lambda^0}$ production and QGSM curves: $\alpha_{SJ}$(0)=0,5 (dashed line)
 and $\alpha_{SJ}$(0)=0,9 (solid line).}
\end{figure}

\end{document}